%% file: 2009DuZa.tex
\documentclass[pre,twocolumn,showpacs]{revtex4}

\include{tex_options}

\newcommand{\leg}{\mrm{leg}}
\newcommand{\self}{\mrm{a}}

\newcommand{\flip}{N}
\newcommand{\Vel}{\bar{\mcal{V}}}
\newcommand{\passive}{\mrm{p}}

\begin{document}
\title{Noisy swimming at low Reynolds numbers}
\author{J\"orn Dunkel}
\email{jorn.dunkel@physics.ox.ac.uk}
\author{Irwin M. Zaid}
\affiliation{Rudolf Peierls Centre for Theoretical Physics, University of Oxford, 1 Keble Road, Oxford OX1 3NP, United Kingdom}
\date{\today}
\begin{abstract}
Small organisms (e.g., bacteria) and artificial microswimmers move due to a combination of active swimming and passive Brownian motion. Considering a simplified linear three-sphere swimmer,  we study how the swimmer size regulates the interplay between self-driven and diffusive behavior at low Reynolds number.  Starting from the Kirkwood-Smoluchowski equation and its corresponding Langevin equation, we derive formulas for the orientation correlation time,  the mean velocity and the mean square displacement in three space dimensions. The validity of the analytical results is illustrated through numerical simulations. Tuning the swimmer parameters to values that are typical of bacteria, we find three characteristic regimes: (i) Brownian motion
at small times, (ii) quasi-ballistic behavior at intermediate time scales, and (iii) quasi-diffusive behavior at large times due to noise-induced rotation.  Our analytical results can be useful for a better quantitative understanding of optimal foraging strategies in bacterial systems, and they can help to construct more efficient artificial microswimmers in fluctuating fluids.
\end{abstract}

\maketitle

\section{Introduction}
Biological~\cite{2005DrEtAl,2007SoEtAl,2008BeEtAl} and artificial microswimmers~\cite{2005DrEtAl,2009LeEtAl,2008TiEtAl,2008TiEtAl_PRL,2007HoEtAl} move through a fluid by performing a series of self-induced shape changes~\cite{1989ShWi,1996StSa}. Handicapped by their tiny size (typically  a few micrometers for a bacterium~\cite{2009CoWe}),  they are forced to swim at very low Reynolds numbers $\Rey\ll 1$~\cite{1883Re,1976Li,1977Pu}. Hence, in order to account for the resulting lack of inertia, the swimming strategies of microorganisms are very different from those operative at human length scales. More precisely, since the fluid flow is reversible at low Reynolds number,  locomotion  in this regime is possible only if the  swimming stroke violates certain time-reversal symmetries~\cite{1977Pu,1989ShWi,1996KoEhMo,2006Ya,2008LaBa}.
\par
Stimulated by experimental advances~\cite{2000Be,2004DoEtAl,2005DrEtAl,2007SoEtAl,2008BeEtAl,2009LeEtAl}, in recent years considerable progress was achieved in understanding the dynamics of deterministic microswimmers models~\cite{2003BeKoSt,2004NaGo,2007IsEtAl,2007EaEtAl,2007PoAlYe,2008AlYe,2008GoAj,2008YaElGo,2008KeMa,2008UnHOGr}. Yet, comparatively little is known quantitatively about the complex interplay between active self-motion, hydrodynamic interactions, and thermal fluctuations in the surrounding fluid~\cite{2003ErEb,2008BaMa_2,2008TeLo,2009BaKuRa}. Very recently, first steps towards clarifying these issues were made by Howse et al.~\cite{2007HoEtAl}, who measured in their experiments the mean square displacement of chemically driven colloidal spheres,  and by Lobaskin et al.~\cite{2008LoLoKu}, who studied the Brownian motion of a triangular microswimmer at  intermediate Reynolds numbers~$\Rey\sim 1$ by combining Lattice Boltzmann simulations with a Langevin description of the swimmer in phase space.  In the present paper, we would like to complement these investigations by concentrating on the diffusive properties of mechanically driven microswimmers at low Reynolds numbers  $\Rey\ll 1$. This limit case is most relevant for bacterial motions and allows one to treat diffusive effects within configuration space. 
\par
Specifically,  we will focus on the following questions: How does the size of the swimmer affect its effective mobility in a noisy fluid? Which details govern the transition from quasi-ballistic self-motion to the diffusive regime? 
To shed light on these issues,  we shall consider simplified quasi-linear $p$-sphere swimmers similar to those  proposed by Najafi and Golestanian~\cite{2004NaGo}. More precisely, we will assume that  internal forces, which generate  
the swimming strokes, are mediated by interaction potentials. 
This approach permits us to treat thermal diffusion effects within the Kirkwood-Smoluchowski
scheme, originally developed to describe the diffusion of polymers in a fluctuating medium~\cite{1972MuAg,1976Bi,2009DuLa,2003LiDu}.
\par
Starting from the Kirkwood-Smoluchowski equation (KSE) ensures that hydrodynamic and stochastic forces are consistently 
coupled on the level of the Fokker-Planck description in configuration space~\cite{1972MuAg}. Moreover, as discussed in Sec.~\ref{s:analytics}, the corresponding Langevin equation  can be used to derive closed analytical formulas for the orientation correlation time,  the mean velocity and the mean square displacement of a single swimmer in three space dimensions. Although the analytical and numerical results in this paper refer to the case of a quasi-linear 3-sphere swimmer $(p=3)$, the formalism can be easily generalized to more complex models (e.g., flexible $p$-sphere swimmer chains). Therefore, this approach can be generally very useful for studying Brownian motion effects and hydrodynamic interactions  in active biological systems at low Reynolds numbers. Furthermore, since it is straightforward to implement an external cofinement (e.g., tweezer or lattice potentials), the Kirkwood-Smoluchowski scheme can help to construct and optimize arrays~\cite{2007GuJo} of, e.g., micro-pumps that work efficiently on those scales where thermal fluctuations in fluid become non-neglible.
\par
Thus, purpose and content of the present paper can be summarized as follows:  First, we will discuss a convenient formalism that allows to simulate active microswimming by means of Langevin equations and interaction potentials (Sec.~\ref{s:theory}). Subsequently, we derive analytic results for the diffusion of a single swimmer (Sec.~\ref{s:analytics}), thereby extending recent work of  Golestanian and Ajdari~\cite{2008GoAj} on deterministic swimmers. A thorough analysis of the single-swimmer case is instructive for a number of reasons: (i)~Exact analytical results provide a useful test of numerical simulations, cf. Sec.~\ref{s:numerics}. (ii)~Recently, Leoni et al.~\cite{2009LeEtAl} were able to experimentally realize an individual  3-sphere system similar to those considered here. (iii)~Understanding the noise-induced behavior of a single swimmer  is a prerequisite for understanding complex behavior and pattern formation in, e.g., self-assembling bacterial systems~\cite{2009CoWe}. (iv)~Dynamical calculations as those presented below may provide a "microscopic" justification for purely probabilistic models of bacterial motility~\cite{1975LoDa}. (v) Depending on the swimmer size, we find a rather sharp transition from purely Brownian to quasi-ballistic motions.  From a (bio-)physical perspective, it is remarkable that the transition occurs when the 3-sphere swimmer model is tuned to bacterial parameters. Hence, loosely speaking, exploiting the interplay between Brownian motion and active swimming may indeed represent a useful strategy (not only) in nature.

\section{General theoretical background}
\label{s:theory}
We consider an ensemble of $N$ microswimmers, each  consisting of $p$ spheres. Neglecting inertia, the state of the system at time $t$ is described by a set of coordinates $(\bs X_\ga)=\{X_{(\ga i)}(t)\}$, where $\ga=1,\ldots, pN$ is a sphere index,  and $i=1,2,3$ labels the space dimension. 
Our subsequent analysis rests on the assumption that the stochastic dynamics of the swimmers in the fluid can be described, at least approximately, by the Kirkwood-Smoluchowski equation (KSE)~\cite{1972MuAg,1976Bi,2009DuLa,2003LiDu}, representing evolution for the $N$-particle probability  density $f(t,\{x_{(\ga i)}\})$.  We begin by recalling  how the KSE can be translated into a Langevin equation for numerical simulations~\cite{2009DuLa,2003LiDu}. Details of the swimming mechanism 
will be discussed in Sec.~\ref{s:swimming_mechanism}.

\subsection{Kirkwood-Smoluchowski and Langevin equation}
Considering a fluid of viscosity $\mu$ and temperature $\Temp$, the KSE reads~\cite{1972MuAg} 
\be\label{e:kirkwood}
\p_t f
=
\p_{(\ga i)} \Hyd_{(\ga i)(\gb j)}
\left\{
\left[\p_{(\gb j)} U \right] f +
\kBT\;\p_{(\gb j)} f
\right\}.
\ee
Here, $\kB$ denotes the Boltzmann constant,  $\p_{(\ga i)}:=\p/\p x_{(\ga i)}$, and a sum is performed over equal double indices $(\ga i)$. The (effective~\cite{1972MuAg}) potential $U$ governs all the internal and external swimmer interactions (see examples in Sec.~\ref{s:swimming_mechanism}), apart from the hydrodynamic interactions mediated by the fluid. The latter are included in the tensor $ \Hyd$.   Considering spherical particles of radius $a_\ga$, the 'diagonal' components of $\Hyd$ are given by
\bse
\be\label{e:hyd-diagonal}
\Hyd_{(\ga i)(\ga j)}
=
\f{\gd_{ ij}}{\gc_{\ga}},
\qquad
\gc_\ga=6\pi \mu a_\ga,
\ee
where $\gd_{ij}$ denotes the Kronecker symbol, and $\gc_\ga$  is the Stokes friction coefficient.
 The hydrodynamic interactions between different spheres are encoded in the 'off-diagonal' components $\Hyd_{(\ga i)(\gb j)}$, $\ga\ne\gb$. If these hydrodynamic interactions are neglected, corresponding to the (infinitely dilute) limit case $\Hyd_{(\ga i)(\gb j)}=0$, Eq.~\eqref{e:kirkwood} reduces to an  
\lq ordinary\rq~ Smoluchowski equation with a diffusion constant $\mcal{D}_\ga=\kB \Temp/\gc_\ga$ for each sphere. 
\par
Here, we are interested in the effects of hydrodynamic interactions, corresponding to $\Hyd_{(\ga i)(\gb j)}\ne 0$. A simple approximation for $\Hyd_{(\ga i)(\gb j)}$, obtained by solving the Stokes equation for a point-like source, is the Oseen-tensor~\cite{Oseen,HappelBrenner}
\be\label{e:Oseen}
\Hyd_{(\ga i)(\gb j)}^O
=
\f{1}{8\pi\mu\, r_{\ga\gb}} 
\biggl(  \gd_{ij} + 
\f{r_{\ga\gb i} r_{\ga\gb j}}{r_{\ga\gb}^2} 
\biggr),
\quad
\ga\ne\gb,
\ee
where $r_{\ga\gb i}:=x_{\ga i}- x_{\gb i}$ and $r_{\ga\gb}:=|\bs x_\ga-\bs x_\gb|$. However, the associated diffusion tensor  $D^O:=\kBT\; \Hyd^O$ is not necessarily positive definite, leading to unphysical behavior if sphere separations become too small~\cite{1968ZwEtAl,1969RoPr}. In our numerical simulations we shall therefore use the improved approximation
\be\notag
\Hyd_{(\ga i)(\gb j)}^M
=
\Hyd_{(\ga i)(\gb j)}^O + 
\f{(a_\ga^2+a_\gb^2)}{24\pi \mu\; r_{\ga\gb}^3}
\biggl(
\gd_{ij} -
3 \f{r_{\ga\gb i} r_{\ga \gb j}}{r_{\ga\gb}^2} 
\biggr)\!,
\\
\label{e:Mazur}
\ee
\ese
which was derived by Mazur~\cite{1982Ma}. The additional term on the right-hand side of Eq.~\eqref{e:Mazur} can be understood
as the next-order correction in a radius-over-distance expansion of the mobility tensor for two spheres~\cite{1982MaSa}. For spheres of equal size ($a_\ga=a_\gb$),
the tensor $\Hyd=\Hyd^M$ defined by Eqs.~\eqref{e:hyd-diagonal} and~\eqref{e:Mazur}, reduces to the Rotne-Prager-Yamakawa tensor~\cite{1969RoPr,1970Ya}. While Eq.~\eqref{e:Mazur} gives a more accurate description than Eq.~\eqref{e:Oseen} at moderate densities, both expressions become invalid if the distance between spheres becomes very small. At very high densities, when sphere-sphere collisions dominate the dynamics,  near-field hydrodynamics and lubrication effects must be modeled more carefully~\cite{2006IsPe}. In the present paper, however, we focus on systems that can be described by Eq.~\eqref{e:Mazur}. 
\par
Unlike the Oseen tensor $\Hyd^O$ from Eq.~\eqref{e:Oseen},  the tensor $\Hyd=\Hyd^M$ is positive definite for $r_{\ga\gb}>a_\ga+a_\gb$ and thus can be  Cholesky-decomposed in the form
\be\label{e:cholesky}
\Hyd_{(\ga i)(\gb j)}=\f{1}{2} \,C_{(\ga i)(\gc k)} C_{(\gb j)(\gc k)}.
\ee
The decomposition~\eqref{e:cholesky} is crucial if one wishes to find a Langevin representation for the stochastic process $\{X_{(\ga i)}(t)\}$ described by the KSE~\eqref{e:kirkwood}:  Upon noting that~\cite{2003WaSzCi,2009DuLa}
\be
\p_{(\ga i)} \Hyd_{(\ga i)(\gb j)}\equiv 0
\ee
holds (for both $\Hyd^O$ and $\Hyd^M$), one finds that the KSE~\eqref{e:kirkwood} corresponds to the following Ito-Langevin equation~\cite{1982HaTh}
\be
\dX_{(\ga i)}(t)
&=&\notag
\Hyd_{(\ga i)(\gb j)} 
F_{(\gb j)}\, \dt +
\\
&& 
(\kB \Temp)^{1/2}
C_{(\ga i)(\gc k)}\, \dB_{(\gc k)}(t).
\label{e:langevin}
\ee
Here, $\{F_{(\gb j)}\}:=\{- \p_{(\gb j)} U\}$ comprises the deterministic forces acting on the spheres, and  $\{B_{(\gc k)}(t)\}$ is a collection of standard Wiener processes; i.e., the increments $\dB_{(\gc k)}(t):=B_{\gc k}(t+\dt)-B_{(\gc k)}(t)$ are independent Gaussian random numbers with distribution
\be
\mcal{P}\{\dB_{(\gc k)}(t) \in [u, u+\diff u]\}=
\f{ e^{ -{u^2}/{(2\,\dt)} }  }
{(2\pi\,\dt)^{1/2}}
\;\diff u,
\ee 
and, according to the Ito scheme~\cite{1982HaTh}, the coefficients $C_{(\ga i)(\gc k)}$ are to be evaluated at time~$t$. For completeness, we still note that, upon formally dividing by $\dt$, the stochastic differential equation~\eqref{e:langevin} can be rewritten as a 'standard' Langevin equation:
\bse
\label{e:langevin_phsyics}
\be
\dot X_{(\ga i)}(t)
&=&\notag
\Hyd_{(\ga i)(\gb j)} F_{(\gb j)}
+\\
&&
(\kB \Temp)^{1/2}C_{(\ga i)(\gc k)}\, \xi_{(\gc k)}(t),
\label{e:langevin-a}
\ee
where $\dot X_{(\ga i)}(t):=\diff X_{(\ga i)}(t)/\dt$ denotes the velocity, and  $\xi_{(\gc k)}(t):=\dB_{(\gc k)}(t)/\dt$ represents Gaussian white noise, i.e.,
\be
\lan \xi_{(\ga i)}(t) \ran &=&0, \\
\lan \xi_{(\ga i)}(t)  \xi_{(\gb j)}(t') \ran &=&\gd_{\ga\gb}\gd_{ij}\; \gd(t-t').
\ee
\ese

\subsection{Swimming mechanism}
\label{s:swimming_mechanism}
Having discussed how to implement fluctuations, we still need to specify the swimming mechanism. To this end, consider two spheres $\ga$ and $\gb$ that form the leg of a swimmer. 
We assume that the internal forces between beads $\ga$ and $\gb$, which generate the swimming stroke, can be derived from a time-dependent potential of the form
\bse\label{e:swimmer_potential}
\be\label{e:swimmer_potential_a}
U_\leg(t,d_{\ga\gb})
=
\f{k_0}{2}\,\{d_{\ga \gb}- [\ell+\gl_{\ga\gb} \sin(\go t +\gvf_{\ga\gb})]\}^2,\;
\ee
with $d_{\ga\gb}(t):=|\bs X_\ga(t)-\bs X_\gb(t)|$  denoting the distance between the spheres, $\gl_{\ga\gb}$ the approximate amplitude of the stroke, and  $\ell > a_\ga+a_{\gb} +\gl_{\ga\gb}$ the mean length of the leg. The potential $U_\leg$ gives rise to two characteristic time-scales: the driving period $T_\go:=2\pi/\go$ and, for a sphere of mass $M_\ga$, the oscillator period $T_{0}:=2\pi/ \sqrt{k_0/M_\ga}$.
Since we are interested in the over-damped regime,  these time scales must be long compared to the characteristic damping time $T_\gc:=M_{\ga}/\gc_\ga$. More precisely, we have  to impose
\be\label{e:timescales}
T_\gc \ll T_0 \ll T_\go\ee
\ese
corresponding to slow driving and fast relaxation. The constraint~\eqref{e:timescales} ensures that our swimmers behave similar to a shape-driven swimmer~\cite{2008GoAj}.
\par
In the remainder, we shall focus on 3-sphere swimmers, representing the smallest self-swimming system within our approach (two-sphere swimmers can achieve active locomotion only due to collective effects~\cite{2008LaBa}). 
We consider three spheres $(\ga,\gb,\gc)$ forming a swimmer with central sphere~$\gb$, e.g., in the case of a single swimmer $(\ga,\gb,\gc)=(1,2,3)$ with middle sphere $\gb=2$. The legs are given by $\bs d_{\ga\gb}:=\bs X_\gb-\bs X_\ga$ and $\bs d_{\gb\gc}:=\bs X_\gc-\bs X_\gb$, and we still define normalized connectors
$$
\bs n_{\ga\gb}:=\bs d_{\ga \gb}/d_{\ga\gb}
\csp
\bs n_{\gb\gc}:=\bs d_{\gb \gc}/d_{\gb\gc}.
$$
In order to ensure that the swimmer moves quasi-linearly~~\cite{2008GoAj},  we introduce a stiffness potential 
\be\label{e:bending_potential}
U_\mrm{lin}
&=&
\f{K}{2}\ell^2\left(
\bs{n}_{\ga\gb}\cdot \bs{n}_{\gb\gc } -1
\right)^2,
\ee
which, for $K\gg k_0$, penalizes bending. The resulting force components $F^\mrm{lin}_{(\ga i)}:=-\p_{(\ga i)} U_\mrm{lin}$ read explicitly
\bse\label{e:bending}
\be
F^\mrm{lin}_{\ga k}
&=&
-\f{Q}{d_{\ga\gb}} ( \gd_{ik} - n_{\ga\gb i} n_{\ga\gb k} ) \,n_{\gb\gc i},
\\
F^\mrm{lin}_{\gc k}
&=&
\f{Q}{d_{\gb\gc}} ( \gd_{ik} - n_{\gb\gc i} n_{\gb\gc k}  ) \,n_{\ga\gb i},\\
F^\mrm{lin}_{\gb k}
&=&
-( F^\mrm{lin}_{\ga k}+F^\mrm{lin}_{\gc k})
\ee
where $n_{\ga\gb i}:=d_{\ga\gb i}/d_{\ga\gb}$, and
\be
Q:
=
K \ell^2
\left(
\bs{n}_{\ga\gb} \cdot \bs{n}_{\gb\gc}  - 1
\right).
\ee
\ese
Equations~\eqref{e:swimmer_potential} and~\eqref{e:bending_potential} provide a convenient way of modeling and simulating rigid or flexible $p$-sphere swimmers by means of potentials.  We note that, by construction,  the sum over the internal swimmer forces is zero. The total potential $U$ appearing in the KSE~\eqref{e:kirkwood} is obtained by summing over all effective interaction potentials~\eqref{e:swimmer_potential} and~\eqref{e:bending_potential}.

\section{Analytical results}
\label{s:analytics}

We next summarize formulae for the correlation time of the orientation vector,  the mean swimmer velocity and the spatial mean square displacement of an isolated 3-sphere swimmer. These analytical results can be obtained from the Langevin equation~\eqref{e:langevin} by using the Oseen approximation~$\Hyd\simeq\Hyd^O$, and  their explicit derivation is discussed in the Appendix~\ref{a:calculations}.
\par
A swimmer's motion can be characterized by its geometric center 
\bse
\be
\bs R(t):=\f{1}{3}(\bs X_1 +\bs X_2 +\bs X_3)
\ee
and the orientation vector
\be
\bs N(t):=\f{\bs X_3 -\bs X_1}{|\bs X_3 -\bs X_1|}.
\ee
\ese
We are interested in determining the mean square displacement
\bse
\be
\Diff_R(t):=\lan [\bs R(t) - \bs R(0)]^2 \ran,
\ee
and the correlation function 
\be
\Diff_N(t):=\lan \bs N(t)  \bs N(0) \ran,
\ee
\ese
where the average is taken over fluctuations in the fluid (i.e., over all realizations of the Wiener process). 
\par
It is convenient to discuss $\Diff_N(t)$ first. For a deterministic initial state $\bs N(0)=(N_k(0))$, we can write $\Diff_N(t) = \lan N_k(t)\ran\,   N_k(0)$ with a summation over equal indices. To obtain an analytical formula for $\lan N_k(t)\ran$, we  assume that Eq.~\eqref{e:timescales} holds true and that bending is neglible, $K\gg k_0$. Then the swimmer behaves like a stiff, shape-driven  Najafi-Golestanian~\cite{2004NaGo} swimmer  and we can approximate
\bse\label{e:Golestanian_approx}
\be
\bs d_{12}&:=&\bs X_2-\bs X_1\simeq\bs N\; d_{12},\\
\bs d_{23}&:=&\bs X_3-\bs X_2\simeq\bs N\; d_{23},\\
\bs d_{13}&:=&\bs X_3-\bs X_1\simeq\bs N\; (d_{12}+d_{23}),
\ee 
where, cf. Eq.~\eqref{e:swimmer_potential_a},
\be
d_{12}&=&\ell +\gl_{12} \sin(\go t+\gvf_{12}),\\
d_{23}&=&\ell +\gl_{23} \sin(\go t+\gvf_{23} ).
\ee
\ese
Adopting the Oseen approximation $\Hyd\simeq\Hyd^O$, one can derive from the Langevin equations~\eqref{e:langevin} the following linear evolution equation (see App.~\ref{a:orientation})
\be\label{e:N-ode}
\lan \dot N_k\ran
=
-
\f{\kBT}{2\pi\mu\,d_{13}^3}\left[\f{2}{3}\left(\f{d_{13}}{a_1}+\f{d_{13}}{a_3}\right)-1\right]
\lan N_k\ran,
\ee
where $\lan \dot N_k\ran:=\lan \diff  N_k(t) /\dt\ran$. 
The $1/a_\ga$-parts are contributions to the rotation rate due to noise on the spheres, whereas the $1/d_{13}^{3}$-contribution is a correction due to hydrodynamic interactions.
Equation~\eqref{e:N-ode} can be solved exactly. The solution exhibits an exponentially decaying oscillatory behavior due to the periodicity of the swimming stroke $d_{13}$. However, for $\ell\gg\max\{\gl_{\ga\gb}\}$ it suffices to approximate $d_{13}\simeq 2\ell$, yielding an exponential decay $\Diff_N(t)\simeq \exp(-t/\tau_\flip)$ with orientation correlation time
\be\label{e:flipping_time}
\tau_\flip
\simeq
\left\{
\f{\kBT}{16\pi\mu\,\ell^3}\left[\f{4}{3}\left(\f{\ell}{a_1}+\f{\ell}{a_3}\right)-1\right]
\right\}^{-1}.
\ee

\par
The time parameter $\tau_\flip$ not only determines the temporal correlation of the orientation vector, it also  plays an important role for the dynamics of the geometric center $\bs R(t)$. As shown in App.~\ref{a:geometric_center_velocity}, the mean swimmer velocity  $\lan \dot{\bs  R}(t)\ran$ is governed by the equation
\bse\label{e:velocity}
\be
\lan \dot R_k\ran
=
\f{A_1}{3}\,
\lan F_{(1k)}\ran
+
\f{A_3}{3} \;
\lan F_{(3k)}\ran,
\ee
where
\be
\lan F_{(1k)}\ran
&=&
-\f{B_3\,\lan \dot{d}_{12k} \ran + C\,\lan \dot{d}_{23k} \ran }{B_1 B_3 -C^2},
\\
\lan F_{(3k)}\ran
&=&
\f{C\,\lan \dot{d}_{12k} \ran + B_1\,\lan \dot{d}_{23k} \ran }{B_1 B_3 -C^2},
\ee
are the noise-averaged internal forces on the first and third sphere (the force on the central sphere can be eliminated by virtue of $F_{(1k)}+F_{(2k)}+F_{(3k)}\equiv 0$), respectively, and
\be
\lan \dot{d}_{12k} \ran
&=&
\lan \dot N_k \ran \, d_{12}+
\lan N_k \ran \, \dot{d}_{12},
\\
\lan \dot{d}_{23k} \ran
&=&
\lan \dot N_k \ran \,d_{23}+
\lan N_k \ran\, \dot{d}_{23}
\ee
the mean change of the leg vectors due stochastic rotations and swimming strokes, with abbreviations
\be
A_1&:=&
\f{1}{\gc_1}-\f{1}{\gc_2} +
\f{1}{4\pi\mu} \left(\f{1}{d_{13}}-\f{1}{d_{23}}\right) ,
\\
A_3&:=&
\f{1}{\gc_3}- \f{1}{\gc_2} +\f{1}{4\pi\mu} \left(\f{1}{d_{13}}-\f{1}{d_{12}}\right) ,
\\
B_1&:=&
\f{1}{\gc_1} +\f{1}{\gc_2}   
-
\f{1}{2\pi\mu\, d_{12}}  ,
\\
B_3&:=&
\f{1}{\gc_2}+
\f{1}{\gc_3} -
\f{1}{2\pi\mu\, d_{23}} ,
\\
C&:=&
\f{1}{\gc_2} -
\f{1}{4\pi\mu} \left(\f{1}{d_{12} }+\f{1}{d_{23}} -\f{1}{d_{13}}\right). 
\ee
 \ese
Since the quantities  $\lan N_k \ran$, $\lan \dot N_k \ran$, $d_{\ga\gb}$, and $\dot{d}_{\ga\gb}$ are known,  Eqs.~\eqref{e:velocity} provide a closed analytical result for the mean velocity $\lan \dot R_k\ran$ of a shape-driven swimmer (within the Oseen approximation). In particular, Eqs.~\eqref{e:velocity} generalize the corresponding velocity formulas for a deterministic swimmer,  recently obtained by  Golestanian and Ajdari~\cite{2008GoAj}, to the "noisy swimming" regime.
\par
For realistic swimmer parameters the orientation correlation time $\tau_\flip$ is typically much larger than the driving period $T_\go=2\pi/\go$. In this case, the rather lengthy result~\eqref{e:velocity} can be considerably simplified (see last part of App.~\ref{a:geometric_center_velocity} for details) to read 
\bse\label{e:velocity_simple}
\be\label{e:velocity_simple-a}
\lan \dot{\bs R}(t)\ran= \Vel\,\lan \bs N(t)\ran.
\ee
Here, $\Vel$ denotes the stroke-averaged velocity (i.e, over an interval $[t,t+T_\go]$) of the corresponding \emph{deterministic} swimmer~\cite{2008GoAj}. By using the approximation~\eqref{e:velocity_simple-a} instead  of the exact results~\eqref{e:velocity} one neglects mean velocity oscillations on small time scales. For example, when considering equal-sized beads with $a_{\ga}=a$ and $\ell \gg \max\{a,\gl_{\ga\gb}\}$, then
\be\label{e:simplification_concrete}
\Vel=\f{7}{24} a\go\left(\f{\gl_{12} \gl_{23}}{\ell^2}\right)\sin{\Gd \gvf},
\ee
\ese
where $\Gd\gvf:=\gvf_{12}-\gvf_{23}$ is the phase difference of the leg contractions, and higher order terms in $a$ and $\gl_{\ga\gb}$ are neglected.
Integrating Eq.~\eqref{e:velocity_simple} with $\lan \bs N(t)\ran\simeq\bs N(0)\exp(-t/\tau_\flip)$, we obtain for the position mean value of the swimmer the simple approximate result
\be
\lan \bs R(t)\ran \simeq \bs R(0)+ \Vel\,\tau_\flip \left[1-\exp(-t/\tau_\flip)\right] \bs N(0);
\ee 
i.e., in the asymptotic limit $t\to\infty$,
\be
\lan \bs R(\infty) \ran =  \bs R(0)+ \Vel\,\tau_\flip \bs N(0).
\ee
 
\par
Finally, let us still consider the mean square displacement $\Diff_R(t):=\lan [\bs R(t) - \bs R(0)]^2 \ran$ for a stiff, shape-driven 3-sphere swimmer described by Eqs.~\eqref{e:Golestanian_approx}.
As discussed in App.~\ref{a:diffusion}, by starting from the Langevin equation for~$\bs R(t)$, one can show that $\Diff_R(t)$ decomposes into the form
\bse\label{e:diffusion}
\be
\Diff_R(t)=
\Diff_R^\passive(t) +
\Diff_R^\self(t),
\ee
where the first part
\be
\Diff_R^\passive(t)
&=&\notag
\f{1}{9} \f{\kBT}{\pi \mu} \left(
\f{1}{a_1} + \f{1}{a_2} + \f{1}{a_3} \right) t+\\
&&
\f{2}{9} \f{\kBT}{\pi\mu} 
\int_0^t \ds\; \left(
\f{1}{d_{12}} + \f{1}{d_{23}} +\f{1}{d_{13}} 
\right)
\quad
\ee
comprises \emph{passive} Brownian motion contributions due to thermal diffusion of the spheres (first line) and hydrodynamic Oseen interactions between them (second line), while the second part
\be\label{e:diffusion_2}
\Diff_R^\self(t) 
\simeq
\Vel^2 
\int_0^t \ds\, \int_0^t\du\; \lan  \bs N(s) \bs N(u) \ran
\ee
\ese
is the contribution due to \emph{active} self-swimming. Similar to Eq.~\eqref{e:velocity_simple}, the expression~\eqref{e:diffusion_2} is valid if the orientation correlation time is much larger than the stroke period,~$\tau_N\gg T_\go$. Inserting the above result $\lan \bs N(t) \bs N(s)\ran \simeq\exp(-t/\tau_\flip) $ and approximating $d_{12}=d_{23}=d_{13}/2\simeq\ell$, we obtain for the spatial  mean square displacement
\be\label{e:diffusion_formula}
\Diff_R(t)
&\simeq&\notag
\f{1}{9} \f{\kBT}{\pi \mu} \left(
\f{1}{a_1} + \f{1}{a_2} + \f{1}{a_3} +\f{5}{\ell}
\right)t+\\
&&
2\, \Vel^2\,\tau_\flip\; [t+\tau_\flip(e^{-t/ \tau_\flip}-1)].
\ee
The approximate result~\eqref{e:diffusion_formula} provides a coarse-grained stroke-averaged description of the translational diffusion (details of the swimming stroke are encoded in $\Vel$). An analogous formula can be used to describe the diffusion of a spherical, chemically driven microswimmer~\cite{2007HoEtAl}. By virtue of Eq.~\eqref {e:diffusion_formula}, we can readily distinguish three distinct regimes:
\par
(i) For $t\ll \tau_\flip$ we can expand the exponential term to linear order and find
\bse
\be
\Diff_R(t)\simeq \Diff^\passive_R(t),
\ee
i.e., passive Brownian motion dominates on very short time scales.
\par
(ii) For $t \lesssim \tau_\flip$ we need to include  terms quadratic in $t/\tau_\flip$ and obtain
\be
\Diff_R(t)\simeq 
\Diff^\passive_R(t)+
\f{\Vel^2t^2}{\tau_\flip},
\ee
i.e., ballistic motion  can dominate on intermediate time scales provided $\Vel^2/\tau_\flip$ is large enough (cf.  examples below).
\par
(iii) For $t\gg \tau_\flip$, we recover diffusive behavior
\be
\lim_{t\to\infty }\f{\Diff_R(t)}{t}= \f{\Diff^\passive_R(t)}{t}+2\Vel^2\tau_\flip.
\ee
\ese
If the swimmer is constructed such that $2\Vel^2\tau_\flip \gg  {\Diff^\passive_R(t)}/{t}$, then its diffusive behavior  at large times is due to noise-induced rotation (with persistence time $\tau_\flip$).
\par
In a sense, the above results provide a \lq microscopic\rq~ justification for the assumptions made by  Lovely and Dahlquist~\cite{1975LoDa}, who studied purely probabilistic models of bacterial motion. We also note that the asymptotic behavior for $t\gg \tau_\flip$ and $2\Vel^2\tau_\flip \gg  {\Diff^\passive_R(t)}/{t}$ agrees qualitatively with results reported by Lobaskin et al.~\cite{2008LoLoKu} for triangular swimmers in the moderate Reynolds number regime $\Rey\sim1$. In this context, we also mention recent work by Golestanian et al.~\cite{2007GoLiAj}, who discuss similar scaling relations for the diffusion of phoretic swimmers. 
\par
In the remainder, we are going to compare the analytical predictions  with results of computer simulations, based on a direct numerical integration of the  Langevin equations~\eqref{e:langevin} for the spheres. Our main focus is on the transition from the passive Brownian motions to the active swimming regime.

\section{Swimmer tuning \& numerical simulations}
\label{s:numerics}
We simulate a single 3-sphere swimmer described by the interaction potentials~\eqref{e:swimmer_potential} and~\eqref{e:bending_potential} and governed by the Langevin equation~\eqref{e:langevin}.  We consider identical spheres of radius $a_\ga=a$, mass $M_\ga=M$, and equal stroke amplitudes $\gl_{12}=\gl_{23}=\gl$. The density of the spheres  is chosen as $\rho=10^{3}$ kg/m$^3$ (water), and the fluid is water at room temperature [$\mu=10^{-3}$ kg/(ms), $T=300$ K]. By fixing the parameters  of the spring and bending potentials as $k_0=10^{-4}$ kg/s$^2$,  $\go=10^3$ Hz and  $K=10 k_0$, we satisfy the time scale condition~\eqref{e:timescales} while ensuring  that the swimmer behaves approximately stiff and shape-driven. The velocity $\Vel$ of the swimmer is optimized by choosing $\Gd\gvf=\pi/2$, cf. Eq.~\eqref{e:simplification_concrete_sim}. 
\par
We are primarily interested in understanding how a change of the swimmer size may affect the diffusive behavior. We therefore fix
the ratios $\ell/a= \ell/\gl=10$ and only vary the leg length $\ell$ from 1 to 10 $\mu\met$ in our simulations (i.e., the mean swimmer length is $\Gl=2\ell$). Put differently, we scale the swimmer proportionally by varying $\ell$. Having specified all parameters, it is useful to summarize the relevant formulae for our choice:
\bse
\be
\label{e:simplification_concrete_sim}
\Vel
&\simeq&
\f{7}{24} a\go\left(\f{\gl^2}{\ell^2}\right)\sin{\Gd \gvf}
\quad
=0.292\; \f{\ell}{\sec},
\\
\label{e:flipping_time_equal_sim} 
\tau_\flip 
&\simeq&
\f{16\pi\mu\,\ell^3}{\kBT}\left(\f{8\ell}{3a}-1\right)^{-1}
\,
=0.473\;\f{\ell^3\,\sec}{\mu\met^3},
\qquad
\\
\f{\Diff_R^\passive}{t}
&\simeq&
\f{1}{9} \f{\kBT}{\pi \mu} \left(
\f{3}{a}+\f{5}{\ell}
\right)
\quad\quad\,
=
5.127\, \f{\mu\met^3}{\ell\,\sec},
\ee
and 
\be
2\Vel^2\tau_\flip&=& 0.080  \f{\ell^5}{ \mu\met^3 \sec}.
\ee
\ese
It is remarkable that increasing the swimmer size  by one order of magnitude increases the orientation correlation time $\tau_\flip$ by three orders of magnitude.
\begin{figure}[t]
\centering
\includegraphics[width=\linewidth]{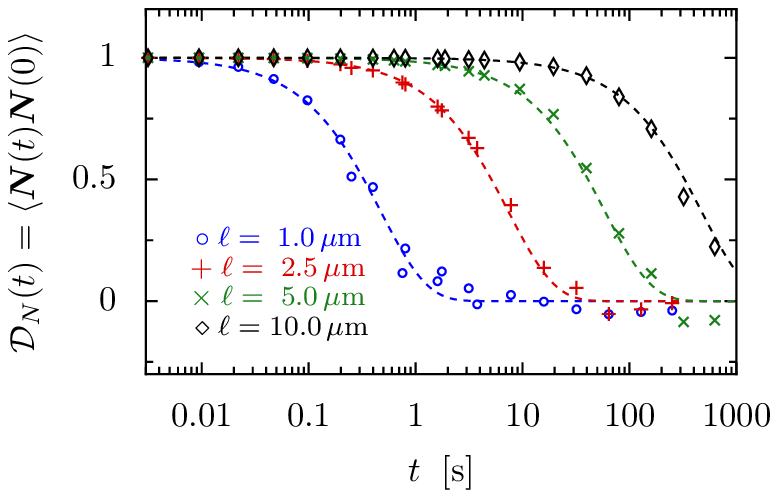}
\caption{
\label{fig:DN}
(Color online) Orientation correlation function for four different swimmer sizes $2\ell$. Symbols  represent averages  over 100 trajectories, numerically calculated from the Langevin equation~\eqref{e:langevin} using parameters as described in the text. The dashed lines  depict the theoretically predicted exponential decay $\Diff_N(t)\simeq \exp(-t/\tau_\flip)$ with orientation correlation time $\tau_\flip$ determined by Eq.~\eqref{e:flipping_time}. It is remarkable that changing the swimmer size by one order of magnitude increases the correlation time by three orders of magnitude.
}
\end{figure}
\begin{figure}[h]
\centering
\includegraphics[width=\linewidth]{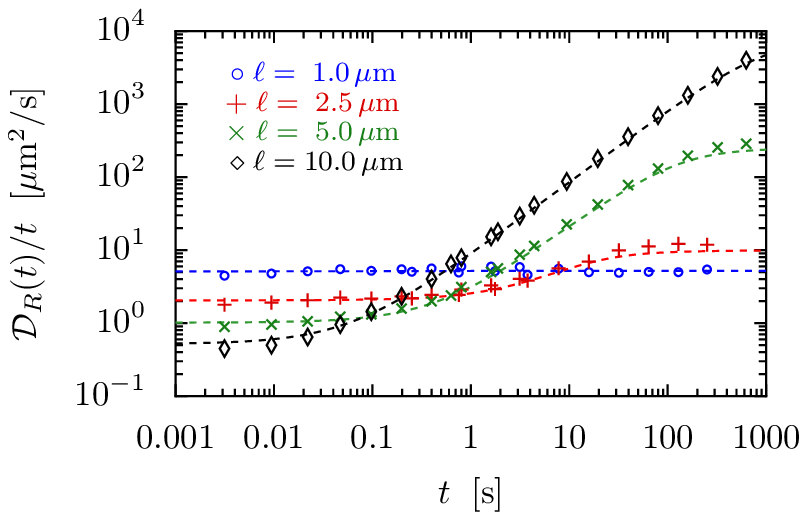}
\caption{
\label{fig:DR}
(Color online) Mean square displacement $\Diff_R(t):=\lan [\bs R(t) - \bs R(0)]^2 \ran$ divided by time $t$ for the same set of parameters as in Fig~\ref{fig:DN}. The dashed lines correspond to the analytical formula~\eqref{e:diffusion_formula}. The dynamics of small swimmers ($\ell=1\;\mu\met$, blue circles) is dominated by Brownian motion on all time scales, whereas big swimmers  ($\ell=10\;\mu\met$, black diamonds) can  move ballistically for several minutes. In the intermediate region ($\ell \sim 5\;\mu\met$, red "$+$"/green "$\times$")  we observe a ballistic transition from ordinary Brownian motion at small times $t\ll\tau_\flip$ to noise-induced rotational diffusion at large times $t\gg\tau_\flip$.   Since $\tau_\flip\sim \ell^3$, the transition from Brownian to quasi-ballistic motion is very sharp. Interestingly enough, typical sizes of bacteria lie in or near this niche~\cite{2009CoWe}.
}
\end{figure}
\par
Figures~\ref{fig:DN} and~\ref{fig:DR} depict the results of numerical simulations (symbols) of the Langevin equations~\eqref{e:langevin} and also the corresponding theoretical predictions (dashed lines). The numerical data points represent averages over 100 trajectories with identical initial conditions. More precisely, at time $t=0$ the swimmer is pointing  along the $x_3$-axis with the first sphere being located at the origin, i.e., $\bs X_1(0)=\bs 0$, $\bs X_2(0)=\ell \bs N(0)$, and  $\bs X_3(0)=[2\ell-\gl \sin(\Gd\gvf)] \bs N(0)$, where $\bs N(0)=(0,0,1)$.
\par
As evident from the diagrams in Fig.~\ref{fig:DN} and~\ref{fig:DR}, the results of the numerical simulations are very well matched by the theoretical curves over several orders of magnitude in time. In particular, for a leg  length in the range $\ell\sim 5\;\mu \met$  (red "$+$"/green "$\times$"-symbols) one readily observes the three aforementioned regimes: (i) Brownian diffusion at small time scales $t\ll \tau_\flip$, (ii) ballistic behavior at intermediate time scales, and (iii) quasi-diffusive behavior due to noise-induced rotation for $t\gg \tau_\flip$.
\par
We conclude the discussion of the numerical results by addressing a few technical aspects  that might be relevant and helpful for future simulations. When considering  ensembles with $N>1$ swimmers the computationally most expensive step is the Cholesky decomposition of the diffusion tensor, see Eq.~\eqref{e:cholesky}, which is approximately of the order $\mcal{O}[(pN)^3]$~\cite{2009DuLa}.  It is also worthwhile to briefly comment on the choice of the time step $\dt$ in the Langevin simulations.  
Ideally, one would like to choose $\dt$  smaller than the smallest dynamical time scale in the system, which for our model is given by the damping time $T_\gc=M/(6\pi\mu a)$. For the swimmer parameters considered here we find $T_\gc\sim 2\times 10^{-9}\;(\ell/\mu\met)^2\sec$, which means that adopting $\dt\sim T_\gc$ would not allow us to simulate experimentally accessible time scales in the seconds range.  Since we are not interested in the dynamical details at very short times scales, we choose in our simulations the time step larger than $T_\gc$, but much smaller than the period $T_\go$ of a swimming stroke by fixing $\dt=10^{-3}T_\go$ for $\ell<5\mu\met$ and  $\dt=10^{-2}T_\go$ if $\ell\ge 5\mu\met$. We verified, however, that  for intermediate time scales (of the order of a few stroke periods $T_\go\sim 6\times 10^{-3}\,\sec$) the numerical results for the mean square displacement and other statistical observables  agree with those obtained for very small time steps $\dt=0.1T_\gc$. Generally, a satisfatcory  resolution of the bending and relaxation dynamics of the legs/spheres would require $\dt \ll (M/K)^{1/2} $ and  $\dt \ll (M/k_0)^{1/2}$, respectively.

\section{Conclusions}
\label{s:conclusions}
Understanding the interplay between Brownian motion, hydrodynamic interactions, and
self-propulsion is a prerequisite for understanding the dynamics of bacteria and
artificial swimming devices at the microscale.  In the first part, we discussed how one can model these phenomena by means of stochastic processes (overdamped Langevin equations). Subsequently, as a first application, we focussed on the size-dependence of diffusive behavior at low Reynolds numbers ($\Rey\ll 1$) for a quasi-linear 3-sphere swimmer model. Our theoretical analysis complements a recent experimental study by Howse et al.~\cite{2007HoEtAl}, who investigated the diffusion of chemically driven, spherical colloids,  and theoretical work by Lobaskin et al.~\cite{2008LoLoKu}, who considered the Brownian dynamics of an artificial triangular microswimmer at moderate Reynolds numbers ($\Rey\sim 1$).
\par
Starting from the Kirkwood-Smoluchowski equation~\cite{1972MuAg,1976Bi,2009DuLa,2003LiDu}, we derived analytical results for the orientation correlation time, the mean velocity, and the mean square displacement of an overdamped, quasi-linear 3-sphere swimmer~\cite{2004NaGo,2008GoAj}.  Analytical formulae as derived here are useful for testing numerical simulations and (in)validating simplified probabilistic models~\cite{1975LoDa}. Moreover, they provide detailed insight into the size-regulated transition from  predominantly random to quasi-ballistic motions. 
\par
The proposed method of modeling swimmers by effective potentials within a Langevin scheme can be readily extended to study complex behavior in larger swimmer ensembles. However, at high swimmer densities,  collisions and near-field hydrodynamics affect the diffusive behavior~\cite{2006IsPe,2007IsPe,2008IsLoPe} and it  will be necessary to modify the hydrodynamic interaction tensor  accordingly. 
Generally, a useful dimensionless  quantifier for the efficiency of active swimming relative to diffusion  is given by the ratio
\be
\eps=\f{\tau_N}{\Lambda/\Vel}.
\ee
The denominator corresponds to the time needed by  a unperturbed swimmer of velocity $\Vel$  to move one body length $\Gl$, and the swimmer geometry is encoded in the orientation correlation time~$\tau_N$. For $\eps\gg 1$ ($\eps\ll 1$) self-propulsion is effective (non-effective). For non-isolated swimmers, $\tau_N$ is not only determined by thermal effects, but also by collisions with other swimmers~\cite{2007IsPe,2008IsLoPe}. 
\par
With regard to future studies we note that the combination of thermal fluctuations and hydrodynamic coupling might also lead
to interesting behavior in simple arrangements of microswimmers. For example, experiments
have shown that colloidal spheres localized in an array of optical traps create memory
effects \cite{1999MeQu} and driven vibrations \cite{2006PoGrQu}. Our formalism provides
a starting point for the investigation of many self-propelled bodies in separate
potentials, which could be helpful for interpreting experimental data of trapped
bacteria or for constructing pumps from a collection of microswimmers.
\par
To summarize, the above results may provide guidance for constructing artificial microswimmers or pumps that work efficiently in the critical transition region that separates Brownian from quasi-deterministic motions. In particular, by tuning the parameters to the narrow cross-over region one could construct swimmers that explore with high probability a maximized volume fraction within a given period of time. The fact that many bacteria live near or exactly in this niche~\cite{2009CoWe} suggests that  this may indeed be a useful strategy. Thus, exploiting the interplay between noise and active self-motion could lead to novel  applications~\cite{2002AsHa}, e.g., with regard to the controlled transport~\cite{2008Go_Cargo} and distribution of chemical and biological substances in small scale technical devices or even within the human body.  

\paragraph*{Acknowledgements}
The authors would like to thank Matthieu Dufay, Bortolo Mognetti,  Olivier Pierre-Louis, Victor Putz, and Julia Yeomans for helpful discussions. This work was supported by the EPSRC grant no. EP/D050952/1 (J.D.).


\appendix
\section{Calculations}
\label{a:calculations}
This appendix provides derivations of analytical results for the orientation correlation function, the mean velocity, and the spatial mean square displacement from the Langevin equation~\eqref{e:langevin}.
Our calculations are based on the following simplifying assumptions: 
\par
(i) The motion of the 3-sphere swimmer is approximately stiff and shape-driven, i.e.,
\bse\label{a-e:stiff}
\be
\bs d_{12}&:=&\bs X_2-\bs X_1\simeq\bs N\; d_{12},\\
\bs d_{23}&:=&\bs X_3-\bs X_2\simeq\bs N\; d_{23},\\
\bs d_{13}&:=&\bs X_3-\bs X_1\simeq\bs N\; (d_{12}+d_{23}),
\ee 
where
\be
d_{12}&=&\ell +\gl_{12} \sin(\go t),\\
d_{23}&=&\ell +\gl_{23} \sin(\go t-\Gd\gvf ).
\ee
\ese
In this case the internal forces, which generate the swimming strokes, point along the swimmer's axis and we may write
\bse
\be
F_{1i}=g_1\,N_i\csp
F_{3i}=g_3\,N_i
\ee
and, with $F_{3i}+F_{2i}+F_{1i}=0$,
\be
F_{2i}=-(g_1+g_3)N_i,
\ee
where $g_\ga$ is the force amplitude, and  $N_i$ denotes a component of the orientation vector
\be
\bs N(t)&:=&\f{\bs X_3 -\bs X_1}{|\bs X_3 -\bs X_1|}.
\ee
\ese
\par
(ii)
We adopt the  Oseen approximation, i.e., $\Hyd\simeq\Hyd^O$ where
\be
\Hyd_{(\ga i)(\gb j)}^O
&:=&\notag
\f{1}{8\pi\mu\, d_{\ga\gb}} 
\biggl(  \gd_{ij} + 
\f{d_{\ga\gb i} d_{\ga\gb j}}{d_{\ga\gb}^2} 
\biggr),
\qquad\\
&\simeq&
\f{1}{8\pi\mu\, d_{\ga\gb}} 
\bigl(  \gd_{ij} + 
N_iN_j 
\bigr).
\label{a-e:Oseen}
\ee
The second line follows from assumption~\eqref{a-e:stiff}. 

\subsection{Orientation correlation function}
\label{a:orientation}
Given a deterministic initial state $\bs N(0)=(N_k(0))$, we are interested in $\Diff_N(t):=\lan \bs N(t)  \bs N(0) \ran = \lan N_k(t)\ran\,   N_k(0)$. 
\par
The first step is to find the stochastic differential equation (SDE) for the orientation vector  $\bs N(t)$. This can be achieved by virtue of the Ito formula~\cite{1982HaTh,2009DuHa}, yielding
\be \label{a-e:langevin_N}
\diff N_k(t)
&=&\notag
\p_{(\ga i)}N_k\, \diff X_{(\ga i)} +\\
&&
D_{(\ga i)(\gb j)} \p_{(\ga i)} \p_{(\gb j)}N_k\, \dt,
\ee
where $\dX_{(\ga i)}(t)$ is governed by Eq.~\eqref{e:langevin}, and the diffusion tensor is given by  $D:=\kBT\;\Hyd\simeq\kBT\;\Hyd^O$.

To determine $ \lan N_k(t)\ran$, we use that  \mbox{$\lan C_{(\ga i)(\gc k)}\, \dB_{(\gc k)}(t) \ran=0$} for an Ito SDE, and therefore
\be\label{a-e:langevin_mean}
\diff \lan X_{(\ga i)(t)}\ran
=
\bigl\lan \Hyd_{(\ga i)(\gb j)} 
{F_{(\gb j)}} \bigr\ran\dt.
\ee
Taking the average of Eq.~\eqref{a-e:langevin_N} and inserting~\eqref{a-e:langevin_mean}, we obtain 
\be
\lan \dot{N}_k(t)\ran
&=&\notag
\left \lan 
\Hyd_{(\ga i)(\gb j)} F_{(\gb j)}  \p_{(\ga i)}N_k\right \ran +\\
&&\left
\lan D_{(\ga i)(\gb j)} \p_{(\ga i)} \p_{(\gb j)}N_k\right\ran,
\label{a-e:langevin_N_mean}
\ee
where $\diff \lan \dot{N}_k(t)\ran:=\lan \diff {N}_k(t)/\dt\ran$. We next evaluate the two terms the rhs.  of Eq.~\eqref{a-e:langevin_N_mean} separately. To this end we note that
\be\label{e-a:derivative}
 \p_{(\ga j)}N_k
=
\f{1}{d_{13}}(\gd_{3\ga}-\gd_{1\ga})
{(\gd_{jk}-N_j N_k)},
\ee 
yielding for the first term 
\be
\Hyd_{(\ga i)(\gb j)} F_{(\gb j)}  \p_{(\ga i)}N_k
&=&\notag
[\Hyd_{(3 i)(\gb j)} - \Hyd_{(1 i)(\gb j)}]\times\\
&&
\f{ F_{(\gb j)} }{d_{13}}(\gd_{ik}-N_i N_k).
\ee
Using Oseen approximation~\eqref{a-e:Oseen}, we find 
\be
\Hyd_{(3 i)(\gb j)} F_{(\gb j)} 
&=&
\left[\f{g_3}{\gc_3}
+
\f{g_1}{4\pi\mu\, d_{31}} -
\f{(g_{3}+g_{1})}{4\pi\mu\, d_{32}} \right]
N_i,
\notag
\ee
and, similarly, $\Hyd_{(1 i)(\gb j)} F_{(\gb j)}\propto N_i$. Taking into account that
\be
 N_i(\gd_{ik}-N_i N_k)\equiv 0,
\ee
the first term on the rhs.  of Eq.~\eqref{a-e:langevin_N_mean} vanishes, and Eq.~\eqref{a-e:langevin_N_mean} reduces to
\be\label{a-e:langevin_N_mean_1}
\lan \dot N_k(t)\ran
&=&
\left\lan D_{(\ga i)(\gb j)} \p_{(\ga i)} \p_{(\gb j)}N_k\right\ran.
\ee
By virtue of Eq.~\eqref{e-a:derivative}, we find
\be
&&\p_{(\ga i)} \p_{(\gb j)}N_k
=\notag
\f{1}{d_{13}^2}(\gd_{3\gb}-\gd_{1\gb})(\gd_{3\ga}-\gd_{1\ga})
\times\\&&
\qquad
\left(3 N_j N_k N_i-N_j \gd_{ik}-N_k \gd_{ij}
- N_i \gd_{jk}
\right).
\qquad\quad
\ee
To obtain the  rhs. of~\eqref{a-e:langevin_N_mean_1}, we still need to contract with the diffusion tensor $D_{(\ga i)(\gb j)}$. This results in two contributions: From the diagonal part we get
\bse
\be
&&\notag
\kBT\sum_{(\ga j)}\f{1}{\gc_\ga}(\gd_{3\ga}-\gd_{1\ga})(\gd_{3\ga}-\gd_{1\ga})
\times\\
&&\notag\quad
\f{1}{d_{13}^2}
\left(3 N_j N_k N_j-N_j \gd_{jk}-N_k \gd_{jj}
- N_j \gd_{jk}
\right)\\
&=&
-\kBT\left(\f{1}{\gc_1}+\f{1}{\gc_3}\right)\f{2N_k}{d_{13}^2},
\ee
while the hydrodynamic off-diagonal terms give
\be
&&\notag
\kBT\sum_{(\ga i)(\gb j)}(1-\gd_{\ga\gb})(\gd_{3\ga}-\gd_{1\ga})(\gd_{3\gb}-\gd_{1\gb})
\times\\
&&\notag\quad
\f{\bigl( \gd_{ij} + N_iN_j \bigr)}{8\pi\mu\, d_{\ga\gb}} 
\times\\
&&\notag\quad
\f{1}{d_{13}^2}
\left(3 N_i N_k N_j-N_j \gd_{ik}-N_k \gd_{ij}
- N_i \gd_{jk}
\right)\\
&=&
\f{\kBT}{2\pi\mu\, d_{13}^3} N_k.
\ee
\ese
Combining the two contributions we find 
\be
\lan \dot N_k(t)\ran
=\notag
\left[-\kBT\left(\f{1}{\gc_1}+\f{1}{\gc_3}\right)\f{2}{d_{13}^2}+\f{\kBT}{2\pi\mu\, d_{13}^3}\right]
\lan N_k\ran,
\ee
which for $\gc_\ga=6\pi\mu a_\ga$  gives Eq.~\eqref{e:N-ode}.

\subsection{Motion of the geometric center}
\label{a:geometric_center}

Following a similar procedure, we can derive analytical expression for the mean velocity $\lan \dot{\bs R}(t)\bs\ran$ and the mean square displacement  $\Diff_{R}(t)$.

\subsubsection{Mean velocity}
\label{a:geometric_center_velocity}

First, we would like to determine the mean velocity $\lan \dot{\bs R}(t)\bs\ran$ of the swimmer's geometric center
\be
\bs R(t)&:=&\f{1}{3}(\bs X_1 +\bs X_2 +\bs X_3).
\ee
Averaging the stochastic differential equation 
\be
\diff \bs R(t)&:=&\f{1}{3}(\diff \bs X_1 +\diff \bs X_2 +\diff \bs X_3)
\ee
with respect to the underlying Wiener process and dividing by $\dt$, we obtain
\be
\lan \dot R_i(t)\bs\ran
&=&\notag
\f{1}{3}\sum_{\ga=1}^3\bigl\lan \Hyd_{(\ga i)(\gb j)} 
F_{(\gb j)}\, \bigr\ran\\
&=&
\f{1}{3}\sum_{\ga=1}^3\biggl\lan 
\f{F_{(\ga i)}}{\gc_\ga}+(1-\gd_{\ga\gb}) \Hyd_{(\ga i)(\gb j)}F_{(\gb j)} \biggr\ran.
\qquad\quad
\label{a-e:velocity_1}
\ee
Considering as before  a stiff, shape-driven swimmer and Oseen interactions $\Hyd=\Hyd^O$, we have $F_{(\gb i)}\simeq g_\gb N_i$ and, therefore,
\be
(1-\gd_{\ga\gb}) \Hyd_{(\ga i)(\gb j)}F_{(\gb j)} 
 &\simeq&\notag
 \sum_\gb
\f{\bigl(  \gd_{ij} + 
N_iN_j 
\bigr)}{8\pi\mu\, d_{\ga\gb}} 
g_\gb N_j\\
&=&
N_i\sum_\gb
\f{g_\gb}{4\pi\mu\, d_{\ga\gb}} .
\ee
Inserting this into Eq.~\eqref{a-e:velocity_1} gives
\be
\lan \dot{R}_i(t)\ran 
&=&\notag
\f{1}{3}\biggl\lan 
N_i\biggl(
\f{g_1}{\gc_1} +\f{g_2}{\gc_2} + \f{g_3}{\gc_3}+\\
&&\notag\qquad
\f{g_2}{4\pi\mu\, d_{12}} +\f{g_3}{4\pi\mu\, d_{13}} +\\
&&\notag\qquad
\f{g_1}{4\pi\mu\, d_{12}} +\f{g_3}{4\pi\mu\, d_{23}} +\\
&&\notag\qquad
\f{g_1}{4\pi\mu\, d_{13}} +\f{g_2}{4\pi\mu\, d_{23}} 
\biggr)
\biggr\ran
\ee
Since the internal swimming forces sum to zero, we may eliminate $g_2$ by using $g_2=-(g_1+g_3)$ to obtain
\be
\lan \dot{R}_i(t)\ran 
&=&\notag
\f{1}{3}\lan 
N_i g_1\ran
\biggl[
\f{1}{\gc_1}-\f{1}{\gc_2} +
\f{1}{4\pi\mu} \left(\f{1}{d_{13}}-\f{1}{d_{23}}\right) 
\biggr]
+\\
&&\notag
\f{1}{3}
\lan 
N_i g_3
\ran
\biggl[
 \f{1}{\gc_3}- \f{1}{\gc_2} +\f{1}{4\pi\mu} \left(\f{1}{d_{13}}-\f{1}{d_{12}}\right) 
\biggr]
\\
&=:&
\f{A_1}{3}\,
\lan N_i g_1\ran
+
\f{A_3}{3} \;
\lan N_i g_3\ran.
\label{a-e:velocity_2extra}
\ee
\par
Consequently, to find $\lan \dot{R}_i(t)\ran $, we still need to determine the mean forces $\lan F_{(\gb i)} \ran=\lan N_i g_\gb \ran$ on the first and last sphere, $\gb=1,3$. This can be achieved as follows: Equation~\eqref{a-e:stiff} implies that
\bse\label{a-e:velocity_3}
\be
\lan \dot{d}_{12k} \ran
&=&
\lan \dot N_k \ran d_{12}+
\lan N_k \ran \dot{d}_{12},
\\
\lan \dot{d}_{23k} \ran
&=&
\lan \dot N_k \ran d_{23}+
\lan N_k \ran \dot{d}_{23}.
\ee
\ese
On the other hand, from the definition of the vectors $\bs d_{\ga\gb}$ and the Langevin equations for $\diff \bs X_{\ga} $, we have
\bse\label{a-e:velocity_4}
\be
\lan \dot{d}_{12k} \ran
&=&
\bigl\lan
\Hyd_{(2k )(\gb j)} 
{F_{(\gb j)}} 
- \Hyd_{(1k )(\gb j)} 
{F_{(\gb j)}}
\bigr\ran,
\qquad
\\
\lan \dot{d}_{23k} \ran
&=&
\bigl\lan
\Hyd_{(3k )(\gb j)} 
{F_{(\gb j)}}
- \Hyd_{(2k )(\gb j)} 
{F_{(\gb j)}}
\bigr\ran.
\qquad\quad
\ee
\ese
Inserting the explicit expressions for $\Hyd_{(\ga i)(\gb j)} $ and $F_{(\gb j)}$, Eqs.~\eqref{a-e:velocity_4} can be rewritten as
\bse
\be
\lan \dot{d}_{12k} \ran
&=&\notag
-
\biggl[
\f{1}{\gc_1} +\f{1}{\gc_2}   
-
\f{1}{2\pi\mu\, d_{12}}   
\biggr]\bigl\lan N_k g_1 \bigr\ran
-\\
&&
\biggl[
\f{1}{\gc_2} -
\f{1}{4\pi\mu} \left(\f{1}{d_{12} }+\f{1}{d_{23}} -\f{1}{d_{13}}\right) 
\biggr]
\bigl\lan N_k g_3 \bigr\ran
\notag\\
&=:&
-B_{1}\; \bigl\lan N_k g_1 \bigr\ran  -  C\;\bigl\lan N_k g_3 \bigr\ran
\ee
and
\be
\lan \dot{d}_{23k} \ran
&=&\notag
\biggl[
\f{1}{\gc_2} -
\f{1}{4\pi\mu\, } 
\left(
\f{1}{d_{12}} +
\f{1}{d_{23}} -
\f{1}{d_{13}}
\right)
\biggl]
\bigl\lan
N_k g_1
\bigr\ran+ 
\\
&&\notag
\biggl[
\f{1}{\gc_2}+
\f{1}{\gc_3} -
\f{1}{2\pi\mu\, d_{23}} 
\biggl]
\bigl\lan
N_k g_3
\bigr\ran
\\
&=:&
C \, \lan N_k g_1\ran   + B_{3}\, \lan N_k g_3 \ran.
\ee
\ese
Hence, in order to obtain the unknown expectation values $\lan N_k g_1\ran$, we have to solve the linear system 
\bse
\be
\lan \dot{d}_{12k} \ran &=&
-B_{1}\, \bigl\lan N_k g_1 \bigr\ran  -  C\,\bigl\lan N_k g_3 \bigr\ran,
\\
\lan \dot{d}_{23k} \ran &=&
C\,  \lan N_k g_1\ran   + B_{3}\, \lan N_k g_3 \ran,
\ee
\ese
with lhs. given by Eqs.~\eqref{a-e:velocity_4}. This is easily done and we may summarize the result for the mean velocity:
\bse
\be
\lan \dot R_k(t)\ran
=
\f{A_1}{3}\,
\lan N_k g_1\ran
+
\f{A_3}{3} \;
\lan N_k g_3\ran,
\ee
where
\be
\lan N_k g_1 \ran
&=&
-\f{B_3\,\lan \dot{d}_{12k} \ran + C\,\lan \dot{d}_{23k} \ran }{B_1 B_3 -C^2},
\\
\lan N_k g_3 \ran
&=&
\f{C\,\lan \dot{d}_{12k} \ran + B_1\,\lan \dot{d}_{23k} \ran }{B_1 B_3 -C^2},
\ee
with
\be
\lan \dot{d}_{12k} \ran
&=&
\lan \dot N_k \ran \, d_{12}+
\lan N_k \ran \, \dot{d}_{12},
\\
\lan \dot{d}_{23k} \ran
&=&
\lan \dot N_k \ran \,d_{23}+
\lan N_k \ran\, \dot{d}_{23}.
\ee
\ese
Since the quantities  $\lan N_k \ran$, $\lan \dot N_k \ran$, $d_{\ga\gb}$, $\dot{d}_{\ga\gb}$, $A_{\ga}$, $B_{\ga}$ and $C$ are known we have thus obtained a closed analytical result for the mean swimmer velocity within the Oseen approximation. Analogous calculations can be performed for $\Hyd^M$, but do not yield much additional insight (for a single swimmer).

\paragraph*{Additional simplifications}
If the orientation correlation time $\tau_\flip$ is larger than the driving period $T=2\pi/\go$ then
\be
\lan \dot{d}_{12k} \ran
\simeq
\lan N_k \ran \, \dot{d}_{12},
\qquad
\lan \dot{d}_{23k} \ran
\simeq
\lan N_k \ran\, \dot{d}_{23}.
\ee
In this case, we may simplify
\bse\label{a-e:velocity-simplified}
\be
\lan \dot R_k(t)\ran
\simeq
\mcal{V}(t)\;  \lan N_k(t) \ran,
\ee
where
\be
\mcal{V}(t)
&=&\notag
-\f{A_1}{3}\left(
\f{B_3\, \dot{d}_{12}  + C\,\dot{d}_{23} }{B_1 B_3 -C^2}
\right)+
\\&&
\f{A_3}{3}\left(
\f{C\,\dot{d}_{12}  + B_1\,\dot{d}_{23} }{B_1 B_3 -C^2}
\right)
\ee
\ese
is a periodic function, $\mcal{V}(t)=\mcal{V}(t+T_\go)$. Since we assumed $\tau_\flip\gg T_\go$, we can  achieve further simplification by replacing $\mcal{V}(t)$ with its stroke-average
\be\label{a-e:velocity-simplified-2}
\Vel:=\int_t^{t+T_\go}\ds\; \mcal{V}(s),
\ee
so that
\be\label{a-e:simplification}
\lan \dot R_k(t)\ran
\simeq
\Vel\;  \lan N_k(t) \ran.
\ee
For example, when considering equal-sized beads with $a_{\ga}=a$ and $\ell \gg \max\{a,\gl_{12},\gl_{23}\}$, then
\be\notag
\mcal{V}(t)=\f{a \go \gl_{12}}{\ell}\cos(\go t+\gvf_{12})+
\f{a \go \gl_{23}}{\ell}\cos(\go t+\gvf_{23}) + \Vel,
\ee
where
\be\label{a-e:simplification_concrete}
\Vel=\f{7}{24} a\go\left(\f{\gl_{12} \gl_{23}}{\ell^2}\right)\sin{\Gd \gvf}
\ee
and $\Gd\gvf:=\gvf_{12}-\gvf_{23}$, and higher order terms have been neglected.

\subsubsection{Spatial diffusion}
\label{a:diffusion}
Using the result for $\diff R(t)$ from above, we may rewrite the mean square displacement $\Diff_R(t)$ as
\be
\Diff_R(t)
&:=&\notag
\lan [\bs R(t)-\bs R(0)]^2  \ran\\
&=&\notag
\left\lan \int_0^t\diff R_k(u)\int_0^t \diff R_k(s)  \right\ran
\\
&=&\notag
\int_0^t \ds\, \int_0^t\du\; \biggl\lan \left[
\f{A_1}{3}\,N_k g_1
+
\f{A_3}{3} \;N_k g_3
\right]_s
\times\\ 
&&\notag
\qquad\qquad\qquad\;\;
\left[
\f{A_1}{3}\,N_k g_1
+
\f{A_3}{3} \;
N_k g_3\right]_u\biggr\ran  +
\\ 
&&\notag
\f{1}{9}\sum_{\ga,\ga'=1}^3
\int\bigl \lan[ C_{(\ga k)(\gc n)}]_s  \, \diff  B_{(\gc n)}(s) \times 
\\
&&\qquad\qquad\quad\,
[C_{(\ga' k)(\gc' n')}]_u\, \diff B_{(\gc' n')}(u) \bigr\ran
\label{a-e:diffusion_1}.
\ee
Here, we have again used that  \mbox{$\lan \int  f(\bs X_{\ga})\;\dB_{(\gc n)}(t)\ran=0$} holds for Ito integrals. We consider the two remaining integrals in Eq.~\eqref{a-e:diffusion_1} separately, starting with the second one. We find
\be
\Diff_R^\passive(t)
&:=&\notag
\f{1}{9}\sum_{\ga,\ga'=1}^3
\int\bigl \lan[ C_{(\ga k)(\gc n)}]_s  \, \diff  B_{(\gc n)}(s) \times  \\
&&\notag\qquad\qquad\quad\,
[C_{(\ga' k)(\gc' n')}]_u\, \diff B_{(\gc' n')}(u) \bigr\ran
\\
&=&\notag
\f{2}{9} \sum_{\ga,\ga'=1}^3\int_0^t \ds\;\lan D_{(\ga k)(\ga' k)} \ran
\\
&=&\notag
\f{2}{9} \kBT \int_0^t \ds\;\biggl[
\f{3}{\gc_1} + \f{3}{\gc_2} + \f{3}{\gc_3} \biggr]+\\
&&\notag
\f{2}{9} \kBT  \sum_{\ga,\ga'=1}^3 (1-\gd_{\ga\ga'})
\int_0^t \ds\; \Hyd^O_{(\ga k)(\ga' k)} ,
\ee
where
\be
 \Hyd^O_{(\ga k)(\ga' k)} 
 =\f{\gd_{kk}+N_k N_k}{8\pi \mu\,d_{\ga\ga'}}
 =\f{1}{2\pi \mu\,d_{\ga\ga'}}.
\ee
For spherical particles we have $\gc_\ga=6\pi\mu a_\ga$ and, therefore,
\be
\Diff_R^\passive(t)
&=&\notag
\f{1}{9} \f{\kBT}{\pi \mu} \left(
\f{1}{a_1} + \f{1}{a_2} + \f{1}{a_3} \right) t+\\
&&\notag
\f{2}{9} \f{\kBT}{\pi\mu} 
\int_0^t \ds\; \left(
\f{1}{d_{12}} + \f{1}{d_{23}} +\f{1}{d_{13}} 
\right).
\ee
For $\ell \gg\max\{\gl_{\ga\gb}\}$, the integrand in the second line can be approximated by  $5/(2\ell)$ yielding
\be\label{a-e:diffusion_2}
\Diff_R^\passive(t)\simeq
\f{1}{9} \f{\kBT}{\pi \mu} \left(
\f{1}{a_1} + \f{1}{a_2} + \f{1}{a_3} +\f{5}{\ell}
\right)\,t.
\ee
It remains to determine the first (double) integral in Eq.~\eqref{a-e:diffusion_1}, reading
\be
\Diff_R^\self(t)&:=&\notag
\int_0^t \ds\, \int_0^t\du\; \biggl\lan \left[
\f{A_1}{3}\,N_k g_1 +
\f{A_3}{3} \;N_k g_3
\right]_s
\times\\ 
&&\notag\qquad\qquad\qquad\;\;
\left[
\f{A_1}{3}\,N_k g_1
+
\f{A_3}{3} \;
N_k g_3\right]_u\biggr\ran.  
\ee
The subscripts indicate the time arguments in the bracketed expressions, respectively.
Upon recalling that  $F_{(\ga i)}=g_\ga N_i$ is the internal force acting on sphere $\ga$, we see that the contribution  $D_R^\self(t)$   is essentially determined by the force-force correlation functions. However, instead of calculating these correlation functions exactly, we may approximate, for  $\tau_\flip\gg T_\go$, the integrand by [cf. Eqs.~\eqref{a-e:velocity_2extra} and~\eqref{a-e:simplification}]
 \be\notag
\lan \left[\ldots
\right]_s 
\left[\ldots \right]_u\ran
\simeq
\Vel^2\,\lan N_k(s)N_k(u)\ran,
\ee
where $\Vel$ is the stroke-averaged velocity of the corresponding deterministic swimmer, cf. Eqs.~\eqref{a-e:velocity-simplified}-\eqref{a-e:simplification}. Adopting this approximation
we find
\be
\Diff_R^\self(t)
&\simeq&\notag
\Vel^2 
\int_0^t \ds\, \int_0^t\du\; \lan  N_k(s) N_k(u) \ran
\\
&=&\notag
\Vel^2 
\int_0^t \ds\, \int_0^t\du\; \exp(-|u-s|/\tau_\flip)
\\
&=&\notag
2\, \Vel^2\,\tau_\flip\; [t+\tau_\flip(e^{-t/ \tau_\flip}-1)],
\ee
and thus the final result
\be\label{a-e:diffusion_formula}
\Diff_R(t)
&=& \notag
\Diff_R^\passive(t)+\Diff_R^\self(t)\\
&\simeq&\notag
\f{1}{9} \f{\kBT}{\pi \mu} \left(
\f{1}{a_1} + \f{1}{a_2} + \f{1}{a_3} +\f{5}{\ell}
\right)\,t+\\
&&
2\, \Vel^2\,\tau_\flip\; [t+\tau_\flip(e^{-t/ \tau_\flip}-1)].
\ee
The first part represents passive (thermal) diffusion, the second part is due to active swimming (note that $\tau_\flip$ is temperature dependent as well).

\bibliography{S_Cilia,S_Experiments,S_Hydro,S_Julia,S_Particle,S_Reviews,S_BrownianSwimmer,S_Polymer_Hydro,Journals,Hanggi,BrownianMotors}

\end{document}

%% file: tex_options.tex
\usepackage{graphicx}
\usepackage{dcolumn}
\usepackage{epstopdf}
\usepackage{eucal}
\usepackage[dvips]{epsfig}
\usepackage{amssymb}

\usepackage{amsmath,amsthm}

\newcommand{\be}{\begin{eqnarray}}
\newcommand{\ee}{\end{eqnarray}}
\newcommand{\bse}{\begin{subequations}}
\newcommand{\ese}{\end{subequations}}

\newcommand{\bnum}{\begin{enumerate}}
\newcommand{\enum}{\end{enumerate}}

\newcommand{\bit}{\begin{itemize}}
\newcommand{\eit}{\end{itemize}}

\newcommand{\bc}{\begin{cases}}
\newcommand{\ec}{\end{cases}}

\newcommand{\bpm}{\begin{pmatrix}}
\newcommand{\epm}{\end{pmatrix}}

\newcommand{\bvm}{\begin{vmatrix}}
\newcommand{\evm}{\end{vmatrix}}

\newcommand{\bs}{\boldsymbol}

\newcommand{\mcal}{\mathcal}

\newcommand{\mrm}{\mathrm}

\newcommand{\ga}{\alpha}
\newcommand{\gb}{\beta}
\newcommand{\gc}{\gamma}
\newcommand{\gd}{\delta}
\newcommand{\eps}{\epsilon}

\newcommand{\gl}{\lambda}

\newcommand{\go}{\omega}

\newcommand{\Gd}{\Delta}

\newcommand{\Gl}{\Lambda}

\newcommand{\gvf}{\varphi}

\newcommand{\p}{\partial}
\newcommand{\f}{\frac}
\newcommand{\diff}{\mrm{d}}

\newcommand{\lan}{\langle}
\newcommand{\ran}{\rangle}

\newcommand{\kB}{k_B}
\newcommand{\Temp}{\mcal{T}}
\newcommand{\kBT}{k_B\Temp}
\newcommand{\Hyd}{\mcal{H}}
\newcommand{\Diff}{\mcal{D}}

\newcommand{\csp}{\;,\qquad}

\newcommand{\du}{\diff u}

\newcommand{\dX}{\diff X}

\newcommand{\dt}{\diff t}
\newcommand{\ds}{\diff s}
\newcommand{\dB}{\diff B}

\newcommand{\met}{\mrm{m}}

\renewcommand{\sec}{\mrm{s}}

\newcommand{\Rey}{\mcal{R}}
